\documentstyle{mn}
\def\simgt{\mathrel{\hbox{\rlap{\hbox{\lower4pt\hbox{$\sim$}}}\hbox{$>$}}}}

\def\m10{m_{10}}

\def\f100{ f_{100} }
\def\f60{ f_{60} }

\def\spose#1{\hbox to 0pt{#1\hss}}
\def\simlt{\mathrel{\spose{\lower 3pt\hbox{$\mathchar"218$}}
     \raise 2.0pt\hbox{$\mathchar"13C$}}}
\def\simgt{\mathrel{\spose{\lower 3pt\hbox{$\mathchar"218$}}
     \raise 2.0pt\hbox{$\mathchar"13E$}}}

\begin{document}
\title[Large ULIRG Survey]{Towards Completing a Large Area ULIRG Survey
\thanks{Based on observations collected at the European
Southern Observatory, La Silla, Chile}}
\author[D.L. Clements et al.]
{D.L. Clements,$^{1,2}$ W.J. Saunders$^3$
and R.G. McMahon$^4$\\
$^1$European Southern Observatory, Karl-Schwarzschild-Strasse 2,
D-85748 Garching-bei-Munchen, Germany\\
$^2$IAS, Batiment 121, Universite Paris XI, 91045 ORSAY Cedex, France\\
$^3$Institute for Astronomy, Edinburgh University, ROE, Blackford Hill,
Edinburgh\\
$^4$Institute of Astronomy, Madingley Road, Cambridge\\
}

\maketitle

\begin{abstract}
We present the latest results in our efforts to produce a complete,
large sample of Ultraluminous Infrared Galaxies (ULIRGs,
L$_{60}>10^{11.4}L_{\odot}$ for H$_0$=100kms$^{-1}$Mpc$^{-1}$,
q$_0$=0.5) for use in the study of these objects and their evolution.
We have been using an optical/IR colour technique to select those
objects most likely to be ULIRGs. 65 of the 198 candidate ULIRGs that
do not yet have redshifts were observed.  Redshifts were obtained for
51 of these objects, including 8 new ULIRGs and a further three
probable ULIRGs with tentative redshifts. The new ULIRGs include one
object at z=0.44 (00029-1424), the highest in the survey so far, and
at least one new broad-line ULIRG (03156-1706) which appears to lie in
a galaxy cluster. We discuss the properties of these objects and their
local environments, concluding that three of them have high excitation
spectra, and that five show disturbed morphology even in the shallow
acquisition images.  We also present the redshift measurements of the
non-ULIRGs identified in the course of this survey.
\end{abstract}


\section{Introduction}

One of the most spectacular results of the IRAS satellite survey was
the discovery of galaxies with huge luminosities in the far-infrared
(far-IR, $\approx$ 8 - 1000 $\mu$m) (e.g Wright et al, 1984).
The far-IR luminosity
of these objects typically exceeds
their optical and UV luminosity by factors of 2 - 80
(see Sanders \& Mirabel, 1996 and references therein).
These far-IR
galaxies are relatively common in the nearby Universe; at lower
luminosities ( 10$^{9.4}$ - 10$^{10.4}$ L$_{\odot}$ H$_o$ = 100 km
s$^{-1}$ Mpc$^{-1}$) their space density is comparable with that of
Markarian starbursts, whilst at the highest luminosities, $>$
10$^{11.4}$ L$_{\odot}$, they are roughly as common as quasars of
similar power output (Sanders et al 1989, Rowan-Robinson
1995). Objects this powerful have been termed ultraluminous infrared
galaxies, ULIRGs.

There are many unanswered questions regarding these ULIRGs. It is
still unclear just what powers their far-IR emission, especially at the
highest luminosities, or what evolutionary role they play in the
development of galaxies.  Recent results from ISO are suggesting that
starbursts, rather than AGN, may dominate their luminosity (Lutz et al
1996), but this issue is still far from settled (see eg. Lonsdale et
al. 1998). 

One of the main obstacles to
our understanding of ULIRGs is the small number so far uncovered in
the IRAS source catalogues.  For example, only 3
\% of the IRAS Bright Galaxy Sample (Sanders et al 1988) had
luminosities $>$ 10$^{11.4}$ L$_{\odot}$, the defining characteristic
of a ULIRG for our assumed value of H$_0$. The 10 ULIRGs discovered in
the BGS are the best studied of this class of object, but it is only
with larger samples that we may perform the statistical studies
necessary to reveal their range of properties and their true
evolutionary role in the universe. Of the 10 BGS ULIRGs, 9 possess
high ionization spectra, typical of objects powered by an active
nucleus, and all 10 seem to be merging or interacting systems (Sanders
et al 1988). On this basis it has been suggested that ULIRGs are an
early stage in the development of quasars.

Larger flux limited surveys of IRAS galaxies, and lists of ULIRGs
discovered as a result, are gradually becoming available. Recent work
includes Kim et al. (1995) and Veilleux et al. (1995), which includes
extensive analysis of both nuclear and off-nuclear spectroscopy, and
Fisher et al. (1995).  Other ULIRG surveys are discussed in Sanders
and Mirabel (1996) and in Clements et al. (1996a, hereinafter Paper 1,
and 1996b).

The evolutionary role of ULIRGs is also of great interest. It is
still unclear whether the rapid evolution of IRAS galaxies
(e.g. Saunders et al. 1990) parallels the density evolution of the
faint blue population (Broadhurst et al. 1988, Colless et al. 1990) or
the luminosity evolution of radio galaxies and quasars (Boyle et
al. 1988; Dunlop \& Peacock, 1990). Samples of IRAS galaxies based on
the IRAS Point Source Catalogue (PSC) contain too few galaxies at
cosmologically significant redshifts to distinguish between the two
types of evolution. Selection from the deeper Faint Source Survey may
be able to reach the necessary depths. We have thus been conducting a
programme to identify candidate ULIRGs over a large region of sky
($\sim$ 0.7 steradians) using the Faint Source Survey to reach a
limiting flux of 0.3Jy at 60$\mu$m. The first results from this survey
programme were discussed in Paper 1. That paper provided basic data on
91 ULIRGs found in our survey region. The present paper presents the
results of further spectroscopy of candidate ULIRGs in the same area,
including the discovery of 11 new or probable ULIRGs.

The rest of this paper is organised as follows. In the next section we
briefly review the selection techniques for our survey. In section 3
we describe the observations and data reduction. Our results are presented in
Section 4, while in section 5 we discuss these results and the current
status of the survey. In section 6 we draw our conclusions. 

We assume H$_0$ = 100 kms$^{-1}$ Mpc$^{-1}$ and q$_0$ = 0.5 throughout.

\section{Sample Selection}

The targets for the present observations are all part of the large
ULIRG survey first described in Paper 1. A detailed description of our
selection techniques can be found there, but we include a brief summary here.

The basic data for the survey is the IRAS Faint Source Survey V1.2
(Moshir et al. 1992) and UK Schmidt Telescope survey plates covering
the region b$<50^o$ and -32.5$^o$ $<$ DEC $<$ +2.5$^o$. IRAS sources
with galaxy-like colours were identified with optical counterparts on
the scanned plates using a maximum likelihood technique (Sutherland \&
Saunders, 1992; Paper 1). To select ULIRG candidates from this list,
we then examine the optical/far-IR flux ratio. It has been found that the
optical galaxy luminosity function has a rather sharp cut-off at high
luminosities, with almost no galaxy having L$\simgt 4 \times 10^{10}
L_{B_{\odot}}$. The Far-IR luminosity function, in contrast, has no such
cut-off. Any galaxy with an extreme IR luminosity will thus have a
large ratio of optical to infrared flux, $\simgt 20$.

We thus define:

\begin{equation}
R_1   = \f60 {\rm dex} [0.4 (B_J - 14.45)] = {\f60 \over f_{B}} 
\end{equation}
\begin{equation}
R_2  = R_1 { \f60 \over 
   {\rm max} (f_{100}, \f60) } 
\end{equation}

and select sources with either $R_1$ or $R_2 > 10$ as candidate ULIRGs.

There are 479 sources that meet this criterion and are above our
adopted 60$\mu$m flux limit of 0.3 Jy in our survey area. Literature
searches and the observations described in Paper 1 have identified 278
of these sources as galaxies and a further 3 as stars. Of the 278
galaxies, 91 turned out to be ULIRGs, suggesting that we have a
$\sim$33\% success rate for finding ULIRGs with this technique. This
contrasts very well with the $\sim3\%$ of IRAS sources found to be
ULIRGs in simple flux limited samples. The observational targets for
the present paper have been selected from the remaining 198 sources
that have yet to have redshifts measured.

\section{Observations and Data Reduction}

The observations were carried out on the ESO 3.6m telescope using the
EFOSC1 instrument on the nights of 29 -- 31st July 1995. The EFOSC1
instrument uses a 512 by 512 Tektronix CCD and can provide both
imaging observations and, via a slit and grism, spectroscopy (Melnick, 1995).
For the present observations, acquisition images were made using the
standard R band filter, and spectroscopy was obtained using the R300
grism. This provides a spectral range of 5940 -- 9770\AA . The
resolution was measured using the arc lamp to be 16.5\AA . In
imaging mode, the plate scale is 0.6 arcseconds per pixel. Acquisition
images were typically 120s long, while spectra received typical integration
times of 300s.

The data were reduced in the usual way using the IRAF system. Bias
frames were subtracted and then the images were flat fielded, using
sky flats for images and dome flats for the spectra. Spectra were
wavelength calibrated using a Helium-Argon arc lamp. Conditions were
generally non-photometric and with seeing ranging from 1 to 2
arcseconds. The spectroscopic standard LTT7379 was used to flatten
the CCD response function and to provide some correction for the A and B
atmospheric absorption bands.

The most prominent line in almost all ULIRG optical spectra is usually
the H$\alpha$6563 + NII6548/6583 doublet blend. In many cases only
this line and the SII6716/6731 doublet are detected, since strong dust
absorption in these objects suppresses emission lines at shorter
wavelengths.  The redshifts for these objects are thus generally
calculated from these two emission lines only, using Gaussian fits to
the lines. The resolution of the observations are not quite high
enough to allow a full deconvolution of the H$\alpha$-NII blend, but
they are high enough to allow an estimate to be made of the relative
strengths of the H$\alpha$ and NII lines in some cases.  If NII
emission is stronger than H$\alpha$, we classify the object as having a 
high excitation spectrum, usually Seyfert 2 or LINER-like.
A high excitation spectrum can indicate objects containing an active
nucleus.
Our previous observations
(Paper 1) using FOSII at the WHT with a resolution of 25\AA ~could not
achieve this.

\section{Results}

Of the 65 objects observed, redshifts were obtained for 51, including
11 new or probable ULIRGs. Five proved to be stars and 6 had
featureless spectra.  Table 1 presents the basic parameters for the
new ULIRGs discovered in this work, while Appendix A provides
information for the non-ULIRG IRAS galaxies that we have
identified. Figure 1 presents the spectra for the new ULIRGs, while
Figure 2 presents their images.

\begin{table*}
\begin{tabular}{rrrrrr}
Name&RA(1950)&DEC(1950)&F$_{60}$(Jy)&z&log(L$_{60} (L_{\odot})$)\\ \hline
00029-1424&00 02 56.8&-14 24 26.0&0.37&0.440&12.08\\
00285-3140&00 28 34.7&-31 40 52.0&0.39&0.219$^*$&11.46\\
00366-0503&00 36 38.6&-05 03 55.0&0.43&0.40$^*$&12.04\\
00415-0737&00 41 33.5&-07 37 38.0&0.40&0.312&11.79\\
00482-2720&00 48 13.9&-27 20 60.0&1.13&0.129&11.45\\
01098-2754&01 09 53.8&-27 54 17.6&0.64&0.220&11.68\\
02226-1658&02 22 36.3&-16 58 26.0&0.53&0.233$^*$&11.65\\
03156-1706&03 15 40.9&-17 06 02.0&0.34&0.351&11.83\\
21498-2636&21 49 49.1&-26 36 12.7&0.33&0.243&11.48\\
23169-0818&23 16 59.3&-08 18 46.0&0.34&0.248&11.51\\
23569-0341&23 56 59.8&-03 41 55.0&0.35&0.304&11.71\\ \hline
\end{tabular}
\caption{Table of newly discovered ULIRGs. Asterisks indicate
tentative redshifts based on single emission lines.}
\end{table*}

\begin{figure*}
\vspace{8in}
\caption{Spectra of the Newly Identified ULIRGs}
Wavelength is given in \AA , while the flux scale is arbitrary. See Table 1
and Section 4.1 for details of the individual objects.
\end{figure*}

\begin{figure*}
\vspace{8in}
\caption{Images of the Newly Identified ULIRGs}
All images are 1x1 arcminute in size, and were taken in the R
band. Typical integration times are 120s. Seeing was between 1 and 2
arcseconds. See Table 1 and Section 4.1 for details of individual objects.
\end{figure*}

\subsection{Notes on Specific Objects}
We here include details on both the spectrum and image of each of the
new ULIRGs.

{\bf 00029-1424} A somewhat broad line is detected and identified
with H$\alpha$ at z=0.44. There are no matching atmospheric features
at this wavelength so we believe it to be a secure detection. A second
line is marginally detected at a wavelength matching SII at the same
redshift. The velocity width of the H$\alpha$ line is $\sim$2000 kms$^{-1}$,
so we tentatively identify this object as an AGN. 

The optical image of this ULIRG is a barely resolved faint source. No
morphological classification is possible.

{\bf 00285-3140} A line is detected at 7986\AA ~which we tentatively
identify as the H$\alpha$-NII blend. We can say nothing about the
nature of this object given the scant information in the spectrum.
The identification as a ULIRG, must be regarded
as tentative since only a single line is detected.

The image of this object is barely resolved. No morphological
assessment can be made.

{\bf 00366-0503} We tentatively identify a feature in the spectrum of
this object at 9175\AA ~with the H$\alpha$-NII blend. This line is not
associated with any atmospheric features, unlike the one at $\sim$
8800\AA.  This places the object at z=0.4, though this redshift must
be regarded as tentative until further confirmation can be
obtained. It is interesting to note that the spectrum for this object
is quite blue, unlike most others in the present paper.

This ULIRG is barely resolved in the optical image. No morphological
assessment is possible.

{\bf 00415-0737} The H$\alpha$-NII blend is clearly detected in this
object, along with SII. The NII emission appears to be very strong,
indicating a likely high excitation spectrum. There is also a marginal
detection of the OIII5007/4959 doublet, but no sign of H$\beta$,
giving log(OIII/H$\beta$) $>$ 0.6 which
suggests an AGN-like ionising spectrum (see Paper 1).

The image of this galaxy shows a distinct plume or extension to the
SE, suggesting that it is a disturbed system.

{\bf 00482-2720} H$\alpha$ is strongly detected in this galaxy, and
marginally resolved from the NII lines. SII is also detected, but is
partially suppressed by the atmospheric A band at 7700\AA . OI6300 is
also detected. On the basis of the marginally resolved H$\alpha$-NII
blend, this galaxy is classified as HII region-like.

This galaxy has a second nucleus or plume to the SW and a possible
plume to the NE, suggesting a disturbed morphology.

{\bf 01098-2754} The H$\alpha$-NII blend is marginally resolved in this object,
with H$\alpha$ being the dominant line. This suggests an HII region-like
spectrum. SII is also clearly detected.

Imaging shows this ULIRG to be a complex system involving two nuclei and
a tail to the E. The separation of the nuclei is about 6 arcseconds.

{\bf 02226-1658} A line is detetecd at 8087\AA ~which we identify as
the H$\alpha$-NII blend. The only other interpretation for this line is
MgII at z=1.89, which would make this an unusually luminous galaxy. We do
not consider this a reasonable interpretation, and favour the lower redshift
value. This adopted redshift, however, must be regarded as tentative.

The image of this ULIRG is undistinguished.

{\bf 03156-1706} This object has a strong broad H$\alpha$ line with a
velocity width of $\sim$ 4000kms$^{-1}$. H$\beta$, with a
similar width, is also detected, as is OIII5007/4959. The broad-line
nature of this object leads us to classify it as a rare IR luminous
quasar.  Structure in the spectrum redward of OIII might be due to
FeII emission which is often associated with strong far-IR emission in
quasars (Lipari et al. 1993, 1994). The object also appears to have an
unusually blue continuum. 

Imaging shows that this object probably lies in a rich group or galaxy cluster.
We do not detect signs of interaction though. An alternative
interpretation is that this object is being gravitationally lensed by a
foreground group or cluster.

{\bf 21498-2636} H$\alpha$ is detected in this object, as is SII,
though only weakly. An unambiguous assessment of the nature of its
spectrum is not possible since the H$\alpha$ detection is not of
sufficient signal-to-noise, but it is most likely to be HII region-like.

The image of this galaxy shows two plumes or tails to the NW and SW,
suggesting a disturbed morphology. A possible interaction companion lies
6 arcseconds to the SE.

{\bf 23169-0818} The H$\alpha$-NII blend is marginally resolved here,
and the SII and OI lines are also detected. The relative strengths of
H$\alpha$ and NII suggest an HII region-like spectrum.

This ULIRG has a second nucleus or plume to the NW. The separation 
between the main and putative second nucleus is $\sim$ 3 arcseconds.

{\bf 23569-0341} Two lines are detected in this object that match with
H$\alpha$-NII and SII at a redshift of 0.304. The H$\alpha$-NII lines are not
strong enough to draw any conclusions about the nature of the spectrum.

The image of this object is an undistinguished, marginally resolved
galaxy.

\section{Discussion}

\subsection{The ULIRG Population}

The present work adds 11 ULIRGs, of which 3 are provisional
identifications, to the large ULIRG survey that we have undertaken,
bringing the total number of ULIRGs in the survey to 102. The highest
redshift in the survey is now 0.44, and a substantial number of
objects have been added between z=0.3 and 0.4. For those objects where
we can attempt some spectral classification, we find that 3/7 have
high ionisation spectra. These include one object that appears to be a
new IR luminous quasar, another with a somewhat broad H$\alpha$ line,
and another with strong NII and OIII emission. This is
consistent with the result in Paper 1 that at least 35\% of ULIRGs
contain AGNs. Morphological studies of the acquisition images show
that five of the eight objects (62\%), where we have deep enough
images to attempt a classification, are in disturbed systems. This
compares with a value of 91\% found by Clements et al. (1996b) and
Murphy et al. (1996) in larger, deeper imaging studies. However, it
has been shown that signs of interaction can often be missed in
shallow images (Clements
\& Baker, 1996), so 5/8 should probably be regarded as a lower
limit.

03156-1706, the broad line IR luminous quasar that we have discovered,
is probably the most interesting object in the present paper. It is
one of only a small number of IRAS selected quasars and is only the
third confirmed broad-line object in our ULIRG survey (the others are
Mrk1014 and 00275-2859). The possible detection of FeII emission is of
particular interest since it has been claimed that FeII emission is
possibly associated with young quasars, just emerging from a dust
obscured ULIRG phase (Lipari et al, 1993, 1994). This object also has
quite a warm 25$\mu$m-to-60$\mu$m colour ($f_{25}/f_{60}=0.39$),
making it similar to the Lipari et al. FeII emitters. High
signal-to-noise and better resolution spectra will be needed to
confirm an FeII detection, but it is interesting to note that the Fe
II emission is probably somewhat weaker than in the strong/extreme Fe
II emitters discussed by Lipari et al. It thus might be a transition
object in a post-starburst phase where the AGN is making a significant
contribution to the far-IR luminosity.

\subsection{Status of the ULIRG Survey}

We have now observed all but 89 objects in our original sample of 479
sources so there is an overall observational completeness of
81\%. Redshifts have been obtained for 329 objects, while ten of these
objects have turned out to be stars, giving a redshift completeness of
70\%, compared with 60\% in Paper 1. Figure 3 shows the present
cumulative completeness as a function of loudness. We also show in
Figure 4 the fraction of ULIRGs among objects with measured redshifts
as a function of IR loudness (R$_1$ and R$_2$). This demonstrates the
effectiveness of our selection technique. We can also use this diagram
to predict the number and IR loudness of the ULIRGs currently awaiting
discovery in our candidate list. This suggests that we have about 40
ULIRGs still to uncover, the vast majority of which will have R$_1 >
50$.

\begin{figure}
\vspace{3in}
\caption{Completeness as a Cumulative Function of IR Loudness}
The upper lines show observational completeness for objects having an
infrared loudness greater than the value indicated on the x axis, while
the lower lines show the same for redshift completeness. Solid lines
are for R$_1$, dotted for R$_2$. [NOTE: The completeness diagram shown
in Paper 1 is misleading. The present diagram is a better representation of
the actual completeness].
\end{figure}

\begin{figure}
\vspace{3in}
\caption{Fraction of ULIRGs as Function of IR Loudness}
The lines give the fraction of ULIRGs among galaxies with a measured
redshift. The low values shown for low loudness values demonstrates
the effectiveness of our selection technique.  The solid line gives
values for R$_1$, the dotted line for R$_2$. We find that R$_1$, our
primary selection parameter, is somewhat better at selecting ULIRGs
than R$_2$.
\end{figure}

In the present catalogue there are now
30 objects that we classify as 'dull'. These are candidate
identifications for the IRAS sources whose spectra show a
strongly detected continuum but no emission lines. 14 objects
previously classified as dull now have redshifts. Most of these have
weak line/continuum ratios, though a few are identified with objects
other than the highest likelihood identification. They do not appear
to be a cohesive class which might bias estimates of luminosity
function evolution if they are not included.

\section{Conclusions}

Continuing observations to identify ULIRGs in our large area survey
have identified 8 clear new ULIRGs and 3 further probable ULIRGs whose
redshifts are in need of confirmation. One of these objects is a new
IR luminous quasar, while another increases the maximum redshift in
the survey to 0.44. The survey is now $\sim$ 70\% complete in redshift
identifications. The loudness selection criteria seem to be working
very well, especially R$_1$, and we predict that there are about 40
more ULIRGs to find in this survey, most of which will have R$_1 >
50$.
\\ ~\\
{\bf ACKNOWLEDGMENTS}
\\~\\
It is a pleasure to thank the support staff at ESO for their help with
the observations. This research has made use of the NASA/IPAC
Extragalactic Database (NED) which is operated by the Jet Propulsion
Laboratory, California Institute of Technology, under contract with
the National Aeronautics and Space Administration. DLC is supported by
an ESO Fellowship and by the EC TMR Network programme,
FMRX-CT96-0068., WS and RGM are supported by Royal Society
Fellowships. We would like to thank the anonymous referee for helpful
comments.

\newpage

{\bf Appendix A}

We here present the new redshifts obtained for non-ultraluminous objects
in the course of this survey.

\begin{table*}
\begin{tabular}{rrrrrr}
Name&RA(1950)&DEC(1950)&f$_{60}$(Jy)&z&log(L$_{60}/L_{\odot}$)\\ \hline
00016-3056& 00 01 39.70&  -30 56 49.2&0.79&0.065&10.68\\
00107-1850& 00 10 46.83&  -18 50 48.8&0.31&0.096&10.62\\
00113-2830& 00 11 22.97&  -28 30 59.4&0.37&0.179&11.25\\
00264-2348& 00 26 28.79&  -23 48 19.2&0.31&0.065&10.28\\
00353-0641& 00 35 21.64&  -06 41 15.9&4.78&0.023&10.55\\
00363-1732& 00 36 20.36&  -17 32 59.2&0.41&0.147&11.12\\
00395-1743& 00 39 35.45&  -17 43 18.1&0.69&0.136&11.28\\
00408-0833& 00 40 51.20&  -08 33 12.0&0.47&0.086&10.71\\
00472{\small +}0010& 00 47 12.42&  {\small +}00 10 51.1&0.40&0.120&10.93\\
00566-1456& 00 56 41.54&  -14 56 38.5&0.53&0.05&10.28\\
01212-0347& 01 21 15.55&  -03 47 38.3&0.61&0.090&10.86\\
01290-1557& 01 29 02.92&  -15 57 06.7&0.45&0.104&10.85\\
01359-0051& 01 35 59.52&  -00 51 35.9&0.34&0.120&10.86\\
01394-0839& 01 39 29.30&  -08 39 47.2&0.36&0.145&11.05\\
01420-2033& 01 42 04.73&  -20 33 56.7&0.53&0.091&10.81\\
01506-2046& 01 50 41.80&  -20 46 30.6&0.46&0.109&10.91\\
01573-1722& 01 57 18.70&  -17 22 54.0&0.35&0.153&11.09\\
02059-1145& 02 05 59.15&  -11 45 46.4&0.49&0.154&11.24\\
02105-1954& 02 10 35.30&  -19 54 37.0&0.38&0.086&10.61\\
02150-2545& 02 15 00.55&  -25 45 20.2&0.33&0.218&11.38\\
02259-2122& 02 25 59.00&  -21 22 18.4&0.42&0.174&11.28\\
03314-2643& 03 31 24.92&  -26 43 35.4&1.84&0.084&11.28\\
03355-3113& 03 35 43.36&  -31 13 38.7&0.46&0.128&11.05\\
21499-2839& 21 49 56.05&  -28 39 31.1&0.32&0.135&10.94\\
22012-2144& 22 01 15.00&  -21 44 31.9&0.51&0.068&10.53\\
22028-1815& 22 02 52.28&  -18 15 22.5&0.45&0.162&11.25\\
22049-2833& 22 04 56.23&  -28 33 05.3&0.38&0.176&11.25\\
22205-2941& 22 20 30.70&  -29 41 20.1&0.61&0.122&11.13\\
22222-0846& 22 22 12.39&  -08 46 20.0&0.35&0.123&10.89\\
22586{\small +}0205& 22 58 44.98&  {\small +}02 05 17.8&0.34&0.042&9.93\\
23003-2625& 23 00 23.51&  -26 25 43.5&0.51&0.088&10.76\\
23033-3134& 23 03 21.48&  -31 34 07.4&0.32&0.192&11.26\\
23131-0403& 23 13 06.70&  -04 03 31.0&0.32&0.122&10.85\\
23131-2854& 23 13 07.30&  -28 54 57.0&0.38&0.185&11.30\\
23279-2224& 23 27 56.04&  -22 24 18.0&0.31&0.175&11.16\\
23390-0217& 23 39 02.10&  -02 17 44.9&0.30&0.167&11.10\\
23394-0502& 23 39 24.01&  -05 02 06.0&0.36&0.08&10.53\\
23575-2352& 23 57 31.63&  -23 52 24.9&0.31&0.100&10.66\\
23584-2704& 23 58 29.38&  -27 04 53.5&0.33&0.137&10.97\\
23598-1327& 23 59 52.90&  -13 27 23.0&0.31&0.181&11.19\\ \hline
\end{tabular}
\caption{Data for non-ULIRGs discovered in the survey}
\end{table*}

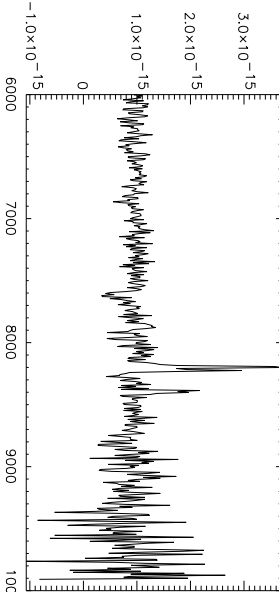
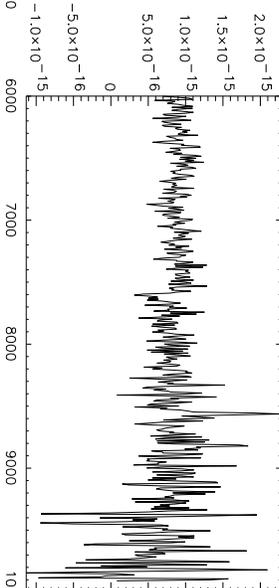
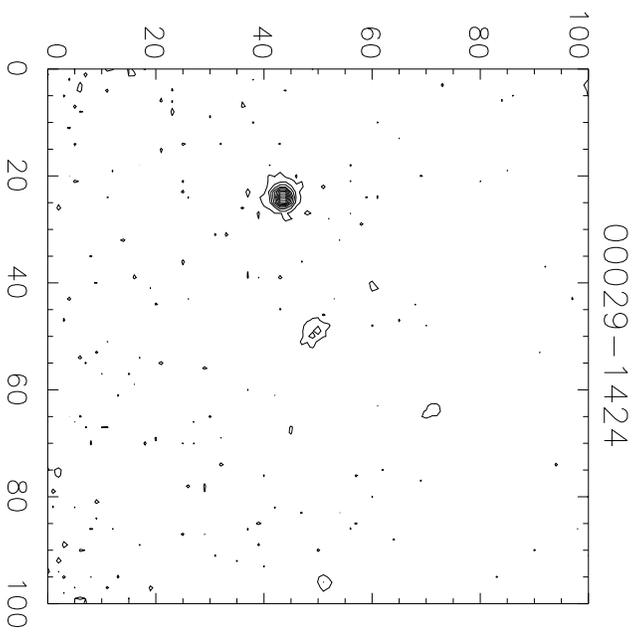
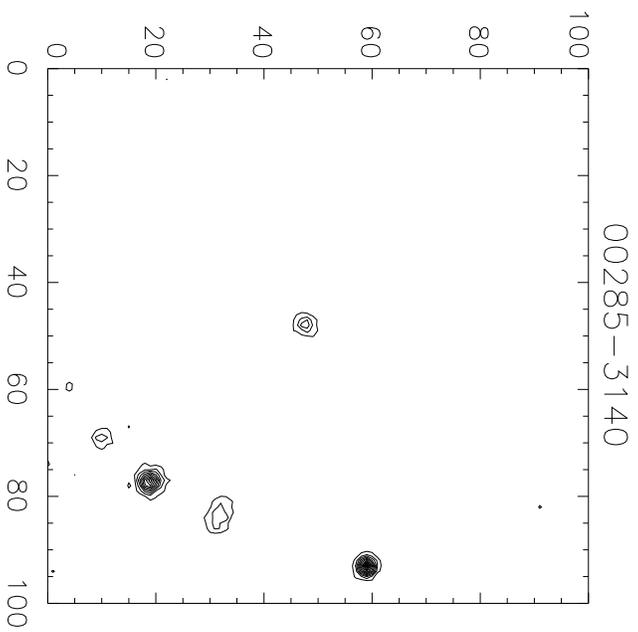
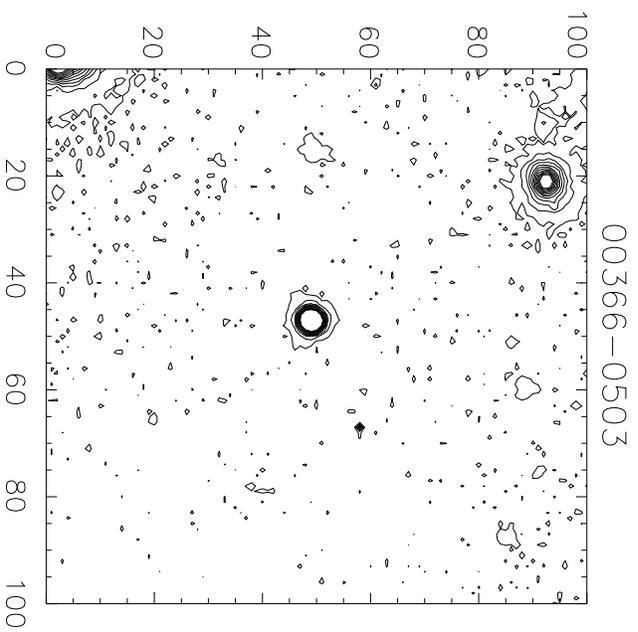
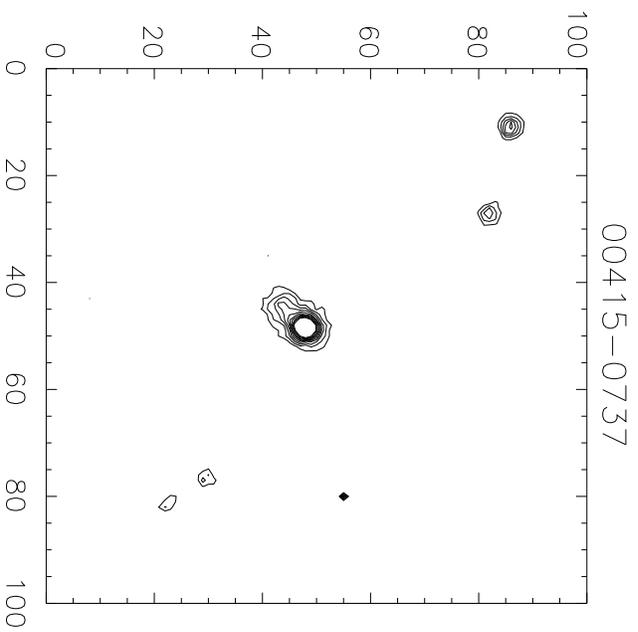
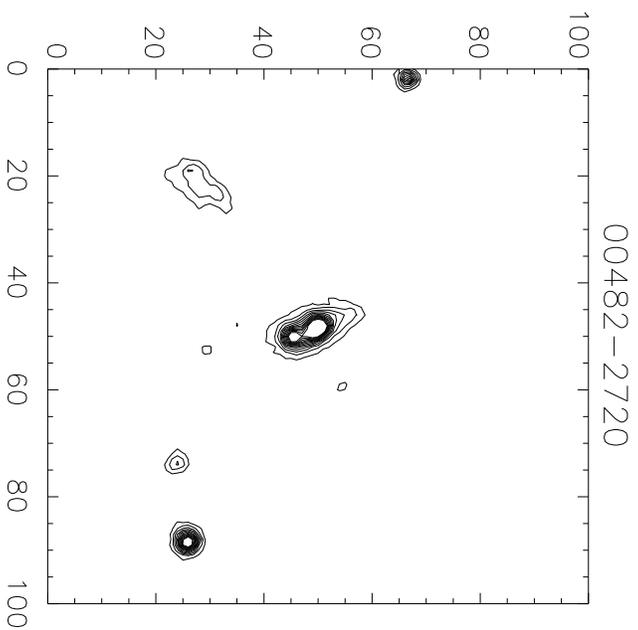
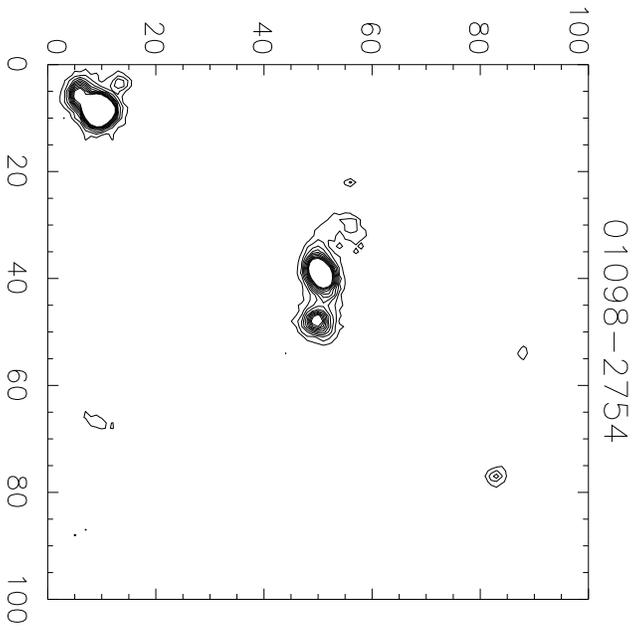
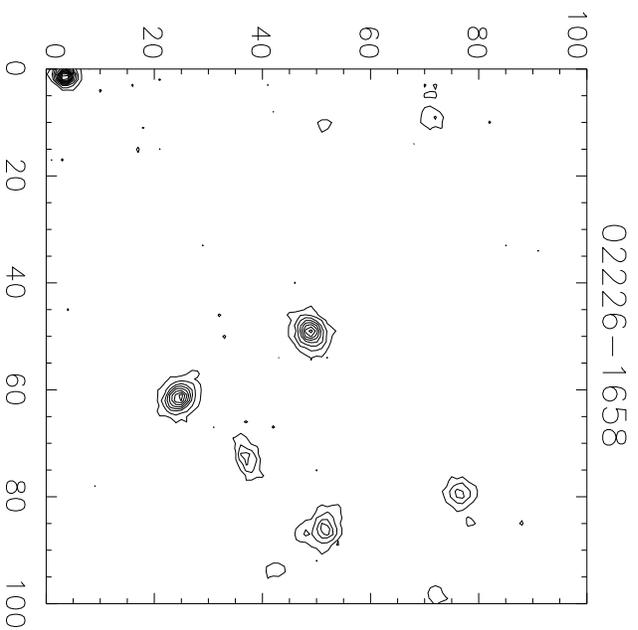
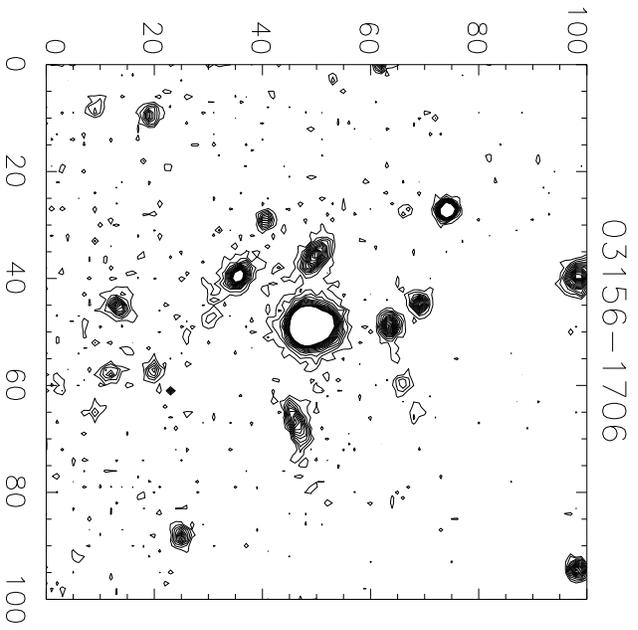
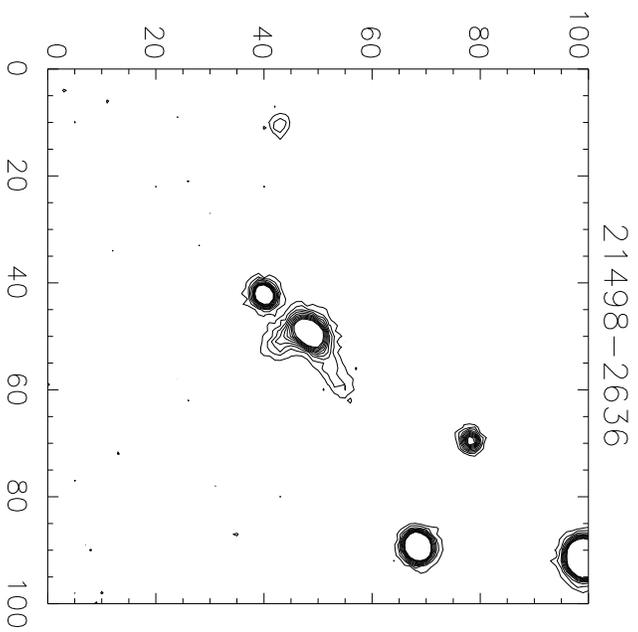
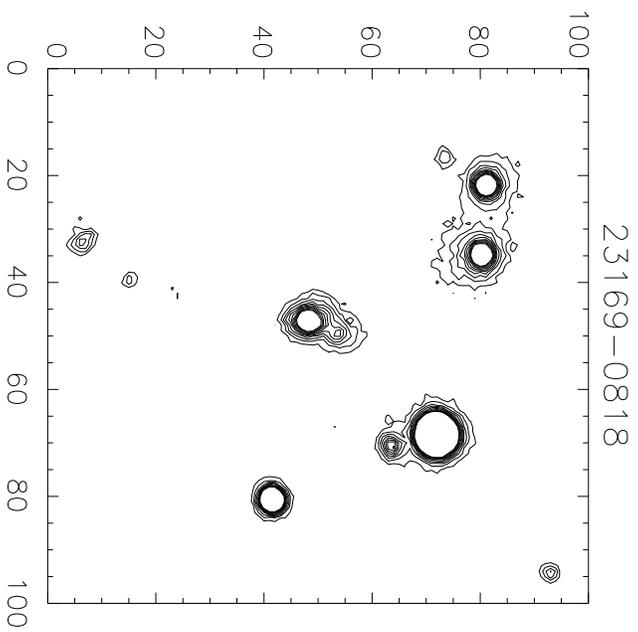
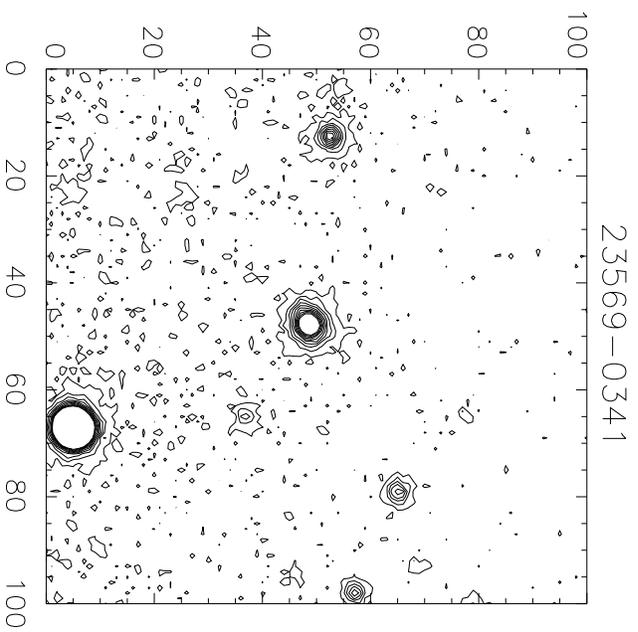
\end{document}